\documentclass{amsart}
\usepackage{amssymb, amsmath, epsfig}
\usepackage{amscd}

\newtheorem{thm}{Theorem}
\newtheorem{prop}[thm]{Proposition}
\newtheorem{lem}[thm]{Lemma}

\newtheorem{remark}{\it Remark}

\newcommand{\R}{\mathbb{ R}}
\newcommand{\h}{\mathfrak{ h}}

\newcommand{\N}{\mathcal{ N}}

\DeclareMathOperator{\pr}{pr} \DeclareMathOperator{\Span}{span}
\DeclareMathOperator{\diag}{diag} 
\DeclareMathOperator{\Ad}{Ad}

\title[Hamiltonization of the Chaplygin Sphere]{Hamiltonization and Integrability of the Chaplygin Sphere in $\R^n$ \footnote{MSC:
37J60, 37J35, 70H45}}
\author{Bo\v zidar Jovanovi\' c}

\begin{document}

\baselineskip=14pt

\maketitle

\centerline{\sc Mathematical Institute SANU} \centerline{\sc Kneza
Mihaila 36, 11000, Belgrade, Serbia} \centerline{{\it e--mail}:
bozaj@mi.sanu.ac.rs}

\begin{abstract}
The paper studies a natural $n$-dimensional generalization of the
classical nonholonomic Chaplygin sphere problem. We prove that for
a specific choice of the inertia operator, the restriction of the
generalized problem onto zero value of the SO(n-1)-momentum
mapping becomes an integrable Hamiltonian system after an
appropriate time reparametrization.
\end{abstract}

\section{Introduction}
Nonholomic systems are not Hamiltonian. Apparently, Chaplygin was
one of the first who considered a time reparametrization in order
to transform nonholonomic systems to the Hamiltonian form
\cite{Ch2}. Also, after \cite{Ch1}, one of the most famous
solvable problems in nonholonomic mechanics, describing the
rolling without slipping of a balanced ball over a horizontal
surface, is referred as the {\it Chaplygin sphere}, see \cite{AKN,
Fe3}. It is interesting that the Hamiltonization of the system by
the use of a time reparametrization was done just recently by
Borisov and Mamaev \cite{BM, BM3} \, (for a geometrical setting
within a framework of almost Poisson brackets, see \cite{GN}).

Fedorov and Kozlov constructed natural $n$-dimensional model of
the Chaplygin-sphere problem  and found an invariant measure
\cite{FeKo}. Various aspects of the problem are studied in
\cite{Sch, Fe2, HN}. In \cite{HN}, it is proved that the reduced
equations of motion of the homogeneous ball are already
Hamiltonian. However, the general problem of integrability and
Hamiltonization is still unsolved.

\subsection{Natural Nonholonomic Systems.} Let $Q$ be a $n$-dimensional Riemannian
manifold with a nondegenerate metric $\kappa(\cdot,\cdot)$, $V:
Q\to \R$ be a smooth function and let $\mathcal D$ be a
nonintegrable $(n-k)$-dimensional distribution of the tangent
bundle $TQ$. A smooth path $q(t)\in Q,\; t\in\Delta$ is called
{\it admissible} (or allowed by constraints) if  the velocity
$\dot q(t)$ belongs to ${\mathcal D}_{q(t)}$ for all $t\in\Delta$.
Let $q=(q_1,\dots,q_n)$ be some local coordinates on $Q$ in which
the constraints are written in the form
\begin{equation}
(\alpha^j_q,\dot q)=\sum_{i=1}^n \alpha_i^j\dot q_i=0,\qquad
j=1,\dots,k, \label{constraints}\end{equation} where $\alpha^j$
are independent 1-forms. The admissible path $q(t)$ is a {\it
motion of the natural mechanical nonholonomic system}
$(Q,\kappa,V,\mathcal D)$ (or a {\it nonholonomic geodesic} for
$V\equiv 0$) if it satisfies the Lagrange-d'Alembert equations
\begin{equation}
\frac{d}{dt}\frac{\partial L}{\partial \dot q_i}+\frac{\partial
L}{\partial q_i}=  \sum_{j=1}^{k}\lambda_j \alpha^j(q)_i,  \qquad
i=1,\dots,n.\label{Hamilton}
\end{equation}

Here the Lagrange multipliers $\lambda_j$ are chosen such that the
solutions $q(t)$ satisfy constraints (\ref{constraints}) and the
Lagrangian is given by the difference of the kinetic and potential
energy: $L(q,\dot q)=\frac12\sum_{ij} \kappa_{ij} \dot q_i \dot
q_j-V(q)$. The expression $\sum_{j=1}^{k}\lambda_j \alpha^j(q)_i$
represents the {\it reaction forces} of the constraints
(\ref{constraints}).

Applying the Legendre transformation $p_i={\partial
L}/\partial\dot q_i=\sum_j \kappa_{ij} \dot q_j$ one can also
write the Lagrange-d'Alembert equations as a first-order system on
the submanifold $\mathcal M =\kappa(\mathcal D)$ of the cotangent
bundle $T^*Q$:
\begin{equation}
 \dot q_i=\frac{\partial
H}{\partial p_i}, \quad \dot p_i=-\frac{\partial H}{\partial q_i}+
\sum_{j=1}^{k}\lambda_j \alpha^j(q)_i, \quad i=1,\dots,n,
\label{ham:eq}
\end{equation}
 where the Hamiltonian is
$H(q,p)=\frac12\sum_{ij}\kappa^{ij}p_ip_j+V(q)$. As for
Hamiltonian systems, it is a first integral of the system.

\subsection{Symmetries, Chaplygin Reduction and Hamiltonization}
Suppose that a Lie group $K$ acts by isometries on $(Q,\kappa)$
preserving the potential function $V$ (the Lagrangian $L$ is $K$-
invariant) and let $\xi_Q$ be the vector field on $Q$ associated
to the action of one-parameter subgroup $\exp(t\xi)$,
$\xi\in\mathfrak k=Lie(K)$. The following version of the {\it
Noether theorem} holds (see \cite{AKN, BKMM}): if $\xi_Q$ is a
section of the distribution $\mathcal D$ then
\begin{equation}
\frac{d}{dt}\left( \frac{\partial L}{\partial \dot q},\,\xi_Q
\right) =\frac{d}{dt}(p,\xi_Q)=0. \label{moment_map}
\end{equation}
In other words, if $\Phi_K: T^*Q \to \mathfrak k^*$ is the
momentum mapping of the $K$-action with respect to the canonical
symplectic structure on $T^*Q$, then $\Phi_K(\xi)$ is conserved
along the flow of \eqref{ham:eq}.

On the other side, suppose that $Q$ has a principal bundle
structure $\pi: Q\to Q/K$ and that $\mathcal D$ is a $K$-invariant
collection of horizontal spaces of a principal connection,
\begin{equation}\label{dim:uslov}
T_q Q= \mathcal D_q\oplus \mathfrak k_q, \quad \mathfrak k_q=\{
\xi_Q(q) \vert \xi\in \mathfrak k \}, \quad q\in Q.
\end{equation}
Then $(Q,\kappa,V, \mathcal D)$ is called a $K$-{\it Chaplygin
system}. The system (\ref{Hamilton}) is $K$-invariant and reduces
to the tangent bundle $T(Q/K)\cong \mathcal D/K$ with the reduced
Lagrangian $L_{red}$ induced from $L\vert_\mathcal D$.

Let $H_{red}$ be a natural mechanical Hamiltonian, the Legendre
transformation of $L_{red}$.
 The reduced vector field $X_{red}$ on the cotangent bundle
$T^*(Q/K)$ can be written in the almost Hamiltonian form
\begin{equation}\label{red:eq}
i_{X_{red}}(\Omega+\Xi)=dH_{red},
\end{equation}
where $\Omega$ is the canonical symplectic form on $T^*(Q/K)$,
$\Xi$ is a semi-basic form depending of the momentum mapping
$\Phi_K$ and the curvature of the connection $\mathcal D$ (for the
details see \cite{Koi, BKMM, CCLM, Tat}).

In some cases the equations (\ref{Hamilton}), i.e, \eqref{ham:eq}
have a rather strong property - an invariant measure (e.g, see
\cite{AKN, BZ}). Within the class of $K$-Chaplygin systems, the
existence of an invariant measure is closely related with their
reduction to a Hamiltonian form.

Suppose that the form $\Omega+\Xi$ is conformally symplectic
$d(\mathcal N(\Omega+\Xi))=0$ (it is assumed that $\N$ is a
function on $Q/K$). In this case the system \eqref{red:eq} has an
invariant measure $\N^{d-1}\Omega^d$, $d=\dim(Q/K)$ and after a
time rescaling $d\tau=\N dt$ it becomes the Hamiltonian system
with respect to the form $\N(\Omega+\Xi)$ . For $d=2$ the above
statement can be inverted: an existence of an invariant measure
implies that the nonholonomic form $\omega+\Xi$ is conformally
symplectic, see \cite{Ch2, St, FeJo, CCLM, Tat, EKR}. The
conformal factor $\mathcal N$ is called the {\it Chaplygin
reducing multiplier}.

Nonholonomic systems on unimodular Lie groups with right-invariant
constraints and left-invariant metrics, so called {\it LR
systems}, always have an invariant measure \cite{VeVe2}. A
nontrivial example of a nonholonomic LR system on the group
$SO(n)$ ({\it $n$-dimensional Veselova problem}), which can be
regarded also as a $SO(n-1)$-Chaplygin system such that the
reduced system on $S^{n-1}=SO(n)/SO(n-1)$ is Hamiltonian after a
time rescaling, is given in \cite{FeJo} (see also Section 5).

The Chaplygin-type reduction and a (partial) Hamiltonization can
be performed also for a class of $K$-invariant noholonomic systems
$(Q,\kappa,V, \mathcal D)$, where the condition \eqref{dim:uslov}
is not satisfied on some $K$-invariant subvariety $S\subset Q$
(see \cite{FeJo2}).

\subsection{Chaplygin Sphere and Reduction of Internal
Symmetries} The $n$-dimensional Chaplygin sphere describes the
rolling without slipping of an $n$-di\-me\-nsi\-onal balanced ball
on an $(n-1)$-dimensional hyperspace $\mathcal H$ in ${\mathbb
R}^{n}$  (\cite{FeKo}, see Section 2 below). This is an
$\R^{n-1}$-Chaplygin system: the kinetic energy and the
nonholonomic distribution $\mathcal D$ are invariant with respect
to the translations of the ball over the hyperplane $\mathcal H$.
After $\R^{n-1}$-reduction it becomes the almost Hamiltonian
system \eqref{red:eq} on the cotangent bundle of the orthogonal
group $SO(n)$,
\begin{equation}\begin{array}{c}
\R^{n-1} \longrightarrow \mathcal D  \subset T(SO(n)\times\R^{n-1}) \qquad\qquad\qquad\qquad   \\
\downarrow  \qquad\qquad\qquad\qquad\qquad\qquad\,\,\\
 \mathcal D/\R^{n-1} \cong TSO(n)\cong T^*SO(n)\,.
\end{array}
\label{bundle*}
\end{equation}

The system is additionally invariant with respect to the
$SO(n-1)$-action - rotations of the ball around the vertical
vector $\Gamma$. The associated vector fields $\xi_{SO(n)\times
\R^{n-1}}$ are sections of the connection \eqref{bundle*} and we
have Noether integrals \eqref{moment_map} that descend to the
conservation law $\dot\Phi=0$ of the reduced flow. Here
\begin{equation}
\Phi: T^*SO(n)\to so(n-1)^* \label{mom:map}
\end{equation}
is the equivariant momentum mapping of the $SO(n-1)$-action with
respect to the canonical form $\Omega$ on $T^*SO(n)$.

However, $\Phi$ is not the momentum mapping with respect to the
nonholonomic form $\Omega+\Xi$. Recently, Hochgerner and
Garcia-Naranjo proved that the form $\Xi$ can be {\it truncated}
to the form $\tilde \Xi$, such that $\Phi$ is the momentum mapping
of the $SO(n-1)$-action on $(T^*SO(n),\Omega+\tilde\Xi)$
\cite{HN}. Moreover, the reduced system is almost Hamiltonian with
respect to $\Omega+\tilde\Xi$ as well:
$i_{X_{red}}(\Omega+\tilde\Xi)=dH_{red}$.

As a result,
following the lines of the usual symplectic
reduction, we can use the momentum mapping $\Phi$ to reduce the
system to the almost Hamiltonian system on $(M_\eta,\mathbf
w_\eta)$, where $M_\eta=\Phi^{-1}(\eta)/SO(n-1)_\eta$,
$SO(n-1)_\eta$ is the coadjoint isotropy group of $\eta\in
so(n-1)^*$:
\begin{equation}\label{red:eq2}
i_{X_{red}^\eta}\mathbf w_\eta=dH_{red}^\eta
\end{equation}
(see \cite{HN}). Now $H_{red}^\eta$ is the induced Hamiltonian
function on $M_\eta$.  So, the Chaplygin multiplier method is
still applicable. In particular, if the ball is homogeneous, the
reduced forms $\mathbf w_\eta$ are closed and the reduced systems
\eqref{red:eq2} are Hamiltonian without a time reparametrization.

Let $\mathcal O_\eta$ be the coadjoint orbit of $\eta$. The
reduced space $M_\eta$ is a $\mathcal O_\eta$-bundle over
$T^*S^{n-1}\cong T^*(SO(n)/SO(n-1))$ that can be seen as a
submanifold of the $SO(n-1)$-reduced space $so(n)^*\times
S^{n-1}$:
\begin{equation}\label{sym:red}
\begin{array}{ccccc}
\mathcal O_\eta \longrightarrow M_\eta \cong \Phi^{-1}(\mathcal
O_\eta)/SO(n-1)& \subset & (T^*SO(n))/SO(n-1)\\
 \downarrow & & \|  \\
   T^*S & & so(n)^* \times S^{n-1}
\end{array}.
\end{equation}

We shall consider the simplest but still very interesting and
nontrivial case, when $\eta=0$. Then the manifold $M_0$ is
diffeomorphic to the cotangent bundle of the sphere $S^{n-1}$ and
the reduced form $\mathbf w_0$ is a semi-basic perturbation of the
canonical symplectic form $\omega$ of $T^*S^{n-1}$.

For the sake of simplicity, denote $\mathbf w_0$, $H_{red}^0$,
$X_{red}^0$, by $\mathbf w$, $H$, $X$, respectively.

\subsection{Outline and Results of the Paper}
In Section 2, we recall the equations of motion of the Chaplygin
sphere. The reduction of the system to the cotangent bundle of the
sphere $T^*S^{n-1}$, for a zero value of the $SO(n-1)$-momentum
mapping $\Phi$ is described in Section 3.

The calculation of an invariant measure as well as the time
reparametrization $d\tau=\N dt$ and the reduction of the system to
the Hamiltonian form for a specific choice of an inertia operator
$I$ of the ball is given in Section 4. On the level of forms, this
means that the form $\mathbf w$  is  conformally symplectic:
$d(\mathcal N\mathbf w)=0$. The description of the Hamiltonization
is given in redundant variables, by the use of a Dirac bracket.

We show that the obtained Hamiltonian system is an integrable
geodesic flow. Moreover, as in the 3-dimensional case \cite{Fe1},
the reduced system is closely related to the {\it associated}
nonholonomic Veselova problem (see Section 5). Namely, the reduced
Veselova problem and the reduced Chaplygin sphere problem share
the same toric foliation of $T^*S^{n-1}$.

In the 3-dimensional case, the group $SO(2)$ is Abelian and all
reduced spaces $M_\eta$ are diffeomorphic to $T^*S^2$. After a
remarkable change of variables, Chaplygin transformed the problem
to the case $\eta=0$ \cite{Ch1}. Since for $n>3$ and $\eta\ne 0$
the coadjoint orbits $\mathcal O_{\eta}$ are nontrivial, some
additional efforts are needed for understanding the complete
dynamics of the ball and it rest still unsolved.

\section{Chaplygin Sphere}

\subsection{Kinematics}
Following \cite{FeKo,Fe2},  consider the Chaplygin-sphere problem
of rolling without slipping of an $n$-dimensional balanced ball
(the mass center $C$ coincides with the geometrical center) of
radius $\rho$ on an $(n-1)$-dimensional hyperspace $\mathcal H$ in
${\mathbb R}^{n}$. For the configuration space we take the {direct
product} of Lie groups $SO(n)$ and $\R^n$, where $g\in SO(n)$ is
the rotation matrix of the sphere (mapping a frame attached to the
body to the space frame) and  $r\in {\mathbb R}^n$ is the position
vector of its center $C$ (in the space frame). For a trajectory
$(g(t),r(t))$ define angular velocities of the sphere in the
moving and the fixed frame, and the velocity in the fixed frame by
$$
\omega=g^{-1}\dot g, \qquad \Omega=\dot g g^{-1}, \quad \mathbf
V=\dot r. $$

In what follows we identify $so(n)\cong so(n)^*$ by an invariant
scalar product
\begin{equation}
 \langle X,Y\rangle=-\frac12\mathrm{tr}(XY).
\label{KF}
\end{equation}

Let $I: so(n) \to so(n)^*\cong so(n)$ be the inertia tensor and
$m$ mass of the ball. The Lagrangian of the system is then given
by
\begin{equation}
L=\frac12 \kappa((\dot \omega,\dot{\mathbf V}),(\dot
\omega,\dot{\mathbf V}))=\frac12\langle
I\omega,\omega\rangle+\frac12 m(\mathbf V,\mathbf V),
\label{ch-lagr}
\end{equation}
where $(\cdot,\cdot)$ is the Euclidean scalar product in $\R^n$.

Let $\Gamma\in{\mathbb R}^{n}$ be a {\it vertical } unit vector
(considered in the fixed frame) orthogonal to the hyperplane
$\mathcal H$ and directed from $\mathcal H$ to the center $C$. The
condition for the sphere to role without slipping leads that the
velocity of the contact point is equal to zero:
\begin{equation} \label{ch-constr}
\mathbf V-\rho{\Omega}\Gamma =0 \, .
\end{equation}

The distribution
$$ \mathcal D=\{(g,r,\omega,\mathbf V) \, \vert\, \mathbf V=\rho
\Ad_g(\omega)\Gamma \}
$$ is right ($SO(n)\times
\R^{n}$)-invariant, so the Chaplygin sphere is an example of a
coupled nonholonomic LR system on the direct product $SO(n)\times
\R^n$ (see \cite{Jo}).

If we take the fixed orthonormal base $ E_1,\dots,E_n$ such that
$\Gamma=E_n$, then the constraint (\ref{ch-constr}) takes the form
$ \dot r_i=\mathbf V_i=\rho\Omega_{in}, \, i=1,\dots,n-1, \, \dot
r_n=\mathbf V_n=0,$ where $\Omega_{ij}=\langle \Omega, E_i \wedge
E_j\rangle$ ($X\wedge Y=X\otimes Y-Y\otimes X=XY^T-YX^T$,
$X,Y\in\R^n$). The last constraint is holonomic, and for the
physical motion we take $r_n=\rho$.

From now on we take $SO(n)\times \R^{n-1}$ for the configuration
space of the rolling sphere, where $\R^{n-1}$ is identified with
the affine hyperplane $\rho\Gamma+\mathcal H$. Then the Chaplygin
sphere is an $\R^{n-1}$-Chaplygin system \eqref{bundle*}, where
the reduced Lagrangian reads
\begin{equation}
L_{red}(\omega,g)= \frac12\langle
I\omega,\omega\rangle+\frac{m\rho^2}{2}(\Ad_g(\omega)\Gamma,\Ad_g(\omega)\Gamma)
=: \frac12\langle\kappa_{red}(g)\,\omega,\omega\rangle.
\label{red:lag}\end{equation}

\begin{remark}{\rm
We can also consider the {\it rubber Chaplygin sphere}, defined as
a system (\ref{ch-lagr}), (\ref{ch-constr}) subjected
 to the additional right-invariant constraints
$\Omega_{ij}=0$, $1\le i < j \le n-1$ describing the no-twist
condition at the contact point \cite{EKR, Jo}.
 }\end{remark}

\subsection{Dynamics} From the
constraints (\ref{ch-constr}) we find the form of reaction forces
in the right-trivialization in which the equations
(\ref{Hamilton}) become
\begin{eqnarray}
&&\dot M=-\rho \Lambda \wedge \Gamma,  \label{c1}\\
&&m\dot{\mathbf V}=\Lambda, \label{c2} \\
&&\dot g=\Omega\cdot g,\label{c3}\\
&& \dot r=\mathbf V.     \label{c4}
\end{eqnarray}
where $M=\Ad_g(I\omega)\in so(n)^*\cong so(n)$ is the ball angular
momentum in the space and  $\Lambda\in\R^n$ is the Lagrange
multiplier.

Differentiating the constraints (\ref{ch-constr}) and using
(\ref{c2}) we get $ \Lambda= m\rho \dot\Omega\Gamma.$ On the other
hand
\begin{equation}
\Lambda \wedge \Gamma=m\rho(\dot\Omega\Gamma)\wedge
\Gamma=m\rho\left( \dot\Omega\, \Gamma \otimes \Gamma + \Gamma
\otimes \Gamma \,\dot\Omega\right)=m\rho\pr_\h(\dot\Omega),
\label{ch}\end{equation} where $\h\subset so(n)$ is the linear
subspace
$\h=\R^n\wedge\Gamma$
and $ \pr_\h: so(n)\to\h$, $\pr_\h(\xi)=(\xi\Gamma)\wedge
\Gamma=\xi\Gamma\otimes\Gamma+\Gamma\otimes\Gamma \xi $ is the
orthogonal projection with respect to the scalar product
(\ref{KF}).

Whence,  (\ref{c1}), (\ref{c3}) is a closed system on $TSO(n)$,
representing the Chaplygin reduction of the $\R^{n-1}$-symmetry.
Now we need to write it in the left trivialization of $TSO(n)$.

Let $\gamma=g^{-1}\Gamma$ be the vertical vector in the frame
attached to the ball. Then
\begin{equation}
\Ad_{g^{-1}}(\h)=\R^n \wedge \gamma=:\mathfrak h^\gamma.
\label{hg}\end{equation}

From the identity
\begin{equation}\label{oO}
\dot\omega=\Ad_{g^{-1}}(\dot\Omega)
\end{equation}
 and the
relations (\ref{ch}) and $ \pr_{\h^\gamma}(\xi)=(\xi\cdot
\gamma)\wedge \gamma={\xi\,} \gamma\otimes\gamma+ \gamma \otimes
\gamma{\,\xi} $ we get
$$
I\dot\omega
=[I\omega,\omega]-m\rho^{2}( {\dot \omega \,} \gamma\otimes\gamma+
\gamma \otimes \gamma{\,\dot\omega} ).
$$

Let us denote $m\rho^2$ by $D$ and let
\begin{equation} \label{K}
\mathbf k=I\omega+D\pr_{\mathfrak h^\gamma} \omega=I\omega+D(
{\omega \,} \gamma\otimes\gamma+ \gamma \otimes \gamma{\,\omega}
)\in so(n)^* \cong so(n)
\end{equation}
be the angular momentum of the ball relative to the contact point
(see \cite{FeKo}). Note that $\mathbf k=\kappa_{red}(g) \omega$,
where the reduced metric $\kappa_{red}(g)$ is defined by
\eqref{red:lag}.

By using the Poisson equation
\begin{equation}
 \dot\gamma=-\omega\gamma \label{PE}
\end{equation}
it easily follows $
\frac{d}{dt}(\omega\gamma\otimes\gamma+\gamma\otimes\gamma\omega)
=\dot\omega\gamma\otimes\gamma+\gamma\otimes\gamma\dot\omega+
[\omega\gamma\otimes\gamma+\gamma\otimes\gamma\omega,\omega]. $

Therefore, the reduced Chaplygin sphere equations, in variables
$(\mathbf k, g)$ of the cotangent bundle $T^*SO(n)$ (or in
variables $(\omega, g)$ of the tangent bundle $TSO(n)$) are given
by
\begin{eqnarray}
&&\dot{\mathbf k}=[\mathbf k,\omega], \label{ch-red}\\
&& \dot g=g\cdot \omega,\label{ch-red2}
\end{eqnarray}
while the reduced kinetic energy is $H_{red}(\mathbf k, g)=\frac12
\langle\kappa^{-1}_{red}(g)\, \mathbf k,\mathbf k\rangle=\frac
12\langle\mathbf k,\omega(\mathbf k)\rangle$.

Let $\Omega$ be the canonical symplectic structure on $T^*SO(n)$,
$d=\dim SO(n)$. It follows from \cite{FeKo, FeRCD} that the
reduced flow on $T^*SO(n)$ has an invariant measure
\begin{equation}
\varrho(\gamma)\vert_{\gamma=g^{-1}\Gamma}\,
\Omega^d=1/\sqrt{\det(\kappa_{red}(g))}\,\Omega^d= 1/\sqrt{\det
({I}+D\pr_{\mathfrak
h^\gamma})}\vert_{\gamma=g^{-1}\Gamma}\,\Omega^d. \label{mu}
\end{equation}

The system is additionally {\it left} $SO(n-1)$-invariant where
the action of $SO(n-1)$ is given by the rotations around the
vertical vector $\Gamma$. The closed system (\ref{PE}),
(\ref{ch-red}) in coordinates $(\mathbf k, \gamma)$ represents the
reduction of $SO(n-1)$-symmetry  to
\begin{equation}
so(n)^*\times S^{n-1}\cong (T^*SO(n))/SO(n-1). \label{SO(n-1):red}
\end{equation}

The volume form \eqref{mu} descends to the invariant measure
\begin{equation}
\varrho(\gamma)\,\varOmega_{so(n)^*} \wedge \varOmega_{S^{n-1}}=
1/\sqrt{\det ({I}+D\pr_{\mathfrak h^\gamma})}\varOmega_{so(n)^*}
\wedge \varOmega_{S^{n-1}}, \label{mu*}
\end{equation}
 where $\varOmega_{so(n)^*}$ and $\varOmega_{S^{n-1}}$ are
standard volume forms on $so(n)^*(\mathbf k)$ and
$S^{n-1}(\gamma)$, respectively (see \cite{FeKo, FeRCD}).

\subsection{Classical Chaplygin Sphere}
In the case $n=3$, under the isomorphism between ${\mathbb R}^3$
and $so(3)$
\begin{equation}
\vec X=(X_1,X_2,X_3)\longmapsto X=\left(\begin{matrix}
0 & -X_3 & X_2 \\
X_3 & 0 & -X_1 \\
-X_2 & X_1 & 0
\end{matrix}\right), \label{iso}
\end{equation}
from (\ref{ch-red}) and (\ref{PE}) we obtain the classical
Chaplygin's ball equations
\begin{equation}
\frac{d}{dt}{\vec{\mathbf k}}=\vec{\mathbf k}\times\vec \omega,
\qquad \frac{d}{dt}{\vec \gamma}=\vec \gamma\times\vec\omega,
\label{Chap}\end{equation} where  $\vec{\mathbf k}=I \vec\omega+
D\vec \omega-D (\vec \omega,\vec\gamma)\vec \gamma$ and $I$ is the
inertia operator of the ball. In the space $(\vec{\mathbf k},
\vec\gamma)$ the density of an invariant measure (\ref{mu*}) is
equal to
\begin{equation}
\varrho(\vec\gamma)=1/\sqrt{ \det(I+D\mathbf
I)\left(1-D(\vec\gamma,(I+D\mathbf I)^{-1}\vec\gamma )\right)},
\label{mu-ch}
\end{equation} the expression given by Chaplygin in \cite{Ch1}.
Since the system \eqref{Chap} has four integrals
\begin{equation}
F_1=(\vec{\mathbf k},\vec \gamma), \quad
F_2=(\vec\gamma,\vec\gamma)=1, \quad F_3=\frac12(\vec{\mathbf
k},\vec \omega), \quad F_4=(\vec{\mathbf k},\vec{\mathbf k}),
 \label{cl:int}
\end{equation}
it is integrable by the Euler-Jacobi theorem: the phase space
$\R^6$ is almost everywhere foliated by invariant tori with
quasi-periodic, non-uniform motion \cite{AKN}. The integration in
\cite{Ch1} is divided into the two steps. Firstly, equations
\eqref{Chap} are solved in the case the area integral $F_1$ is
zero, using elliptic coordinates on the Poisson sphere $F_2=1$.
Then, after an ingenious linear change of variables $(\vec{\mathbf
k},\vec{\gamma}) \longmapsto (\vec{\mathbf k}_1,\vec{\gamma}_1)$,
the problem transforms to the zero area case.

\section{Reduced System in Redundant Coordinates}

\subsection{Reduction to $T^*S^{n-1}$}
From (\ref{c1}) we have
\begin{eqnarray}
\frac{d}{dt}(\pr_{so(n-1)} M)&=&\frac{d}{dt}(\pr_{so(n-1)} \Ad_g
(I\omega))\nonumber\\
&=& \frac{d}{dt}(\pr_{so(n-1)} \Ad_g
(I\omega+D\pr_{\h^\gamma}\omega))\label{cl}\\
&=&\frac{d}{dt}(\pr_{so(n-1)} \Ad_g \mathbf k)=  0,\nonumber
\end{eqnarray}
where $so(n-1)\subset so(n)$ is orthogonal complement to
$\h=\R^n\wedge\Gamma$ with respect to the scalar product
(\ref{KF}).

The integral (\ref{cl}) is actually the momentum mapping
\eqref{mom:map} of the left $SO(n-1)$-action. For $n=3$ we have
the classical area integral $F_1=(\vec{\mathbf
k},\vec\gamma)=(I\vec\omega,\vec\gamma)$.

So we can pass to the reduced system \eqref{red:eq2} on
$M_\eta=\Phi^{-1}(\eta)/SO(n-1)_\eta$ (\cite{HN}, see
Introduction). We shall consider the simplest but still very
interesting case, when we fix the value of the momentum mapping
$\Phi$ to be zero
\begin{equation}
\pr_{so(n-1)} \Ad_g (I\omega)=\pr_{so(n-1)} \Ad_g \mathbf
k=\pr_{so(n-1)^\gamma} (I\omega)=\pr_{so(n-1)^\gamma}\mathbf k=0.
\label{zero}
\end{equation}
Here
$so(n-1)^\gamma:=\Ad_{g^{-1}}so(n-1)=(\R^n\wedge\gamma)^\perp=(\h^\gamma)^\perp.
$

Whence, both $\mathbf k$ and $I\omega$ belong to the subspace
(\ref{hg}). Now, let us introduce new variables $p, \,\xi\in\R^n$
orthogonal to $\gamma$
\begin{equation}\label{uslovi}
(\gamma,p)=(\gamma,\xi)=0,
\end{equation}
such that
\begin{equation}
\mathbf k=\gamma \wedge p, \qquad \omega=I^{-1}(\gamma \wedge
\xi).\label{smene}
\end{equation}

\begin{lem}
The variables $p$ and $\xi$ are related via
\begin{equation}
p=\xi-DI^{-1}(\gamma\wedge\xi)\gamma \label{veza}
\end{equation}
\end{lem}

\begin{proof} The proof directly follows from the definition $\mathbf
k=I\omega+D((\omega\gamma)\wedge \gamma)$ and relations
(\ref{smene}). \end{proof}

From (\ref{veza}), under the conditions (\ref{uslovi}), the
variable $\xi$ can be uniquely expressed via $p$ and $\gamma$.

Note that the coordinates $(\gamma,p)$ can be considered as
redundant coordinates of the cotangent bundle of the sphere
$T^*S^{n-1}$ realized as a subvariety of ${\mathbb R}^{2n}$
defined by constraints
\begin{equation}
\phi_1\equiv (\gamma,\gamma)=1, \quad \phi_2\equiv (\gamma,p)=0.
\label{phi}
\end{equation}

\begin{thm}\label{T1}
The reduced Chaplygin-sphere problem  on
$T^*S^{n-1}=\Phi^{-1}(0)/SO(n-1)$ is described by the equations
\begin{eqnarray}
&& \dot \gamma=X_\gamma(\gamma,p)=-\omega \gamma=-I^{-1}(\gamma\wedge \xi(\gamma,p))\gamma\label{r1}\\
&&\dot p=X_p(\gamma,p)=-\omega p=-I^{-1}(\gamma\wedge
\xi(\gamma,p))p\label{r2}
\end{eqnarray}
\end{thm}

\begin{proof}
The mapping $(\gamma,p)\mapsto (\mathbf k=\gamma \wedge p,\gamma)$
realizes $T^*S^{n-1}$ as a submanifold of \eqref{SO(n-1):red} (see
diagram \eqref{sym:red}).  The equation (\ref{r1}) follows
directly from the Poisson equation (\ref{PE}). On the other hand,
from the equation (\ref{ch-red}) we get
\begin{eqnarray*}
&& \dot \gamma \wedge p+\gamma \wedge \dot p  = [\gamma \wedge p,\omega]\\
\Longrightarrow && -\omega \gamma
p^T-p(-\omega\gamma)^T+\gamma\wedge \dot p= \gamma
p^T\omega-p\gamma^T \omega-\omega\gamma p^T + \omega p\gamma^T\\
\Longrightarrow && \gamma \wedge \dot p = \omega p \gamma^T+\gamma
p^T \omega=(\omega p)\wedge \gamma \\
\Longrightarrow && \dot p =-\omega p +\lambda \gamma.
\end{eqnarray*}

The multiplier $\lambda$ is equal to zero. Indeed, from
(\ref{phi}) we have
$$
\frac{d}{dt}\phi_2=(\dot \gamma,p)+(\gamma,\dot p)=(-\omega
\gamma,p)+(\gamma,-\omega p)+\lambda(\gamma,\gamma)=\lambda=0.
$$
\end{proof}

Note that the reduced Hamiltonian
\begin{equation}\label{red:ham}
H(\gamma,p)=\frac12\langle \mathbf k,\omega\rangle= \frac12\langle
\gamma \wedge p, I^{-1}(\gamma \wedge \xi(\gamma,p))\rangle
\end{equation}
(which is now unique only on the subvariety (\ref{phi})) as well
as the system (\ref{r1}), (\ref{r2}) itself, is defined on
\begin{equation}
\hat\R^{2n}=\R^{2n}\setminus\{\gamma=0\}. \label{hat}
\end{equation}

Also considered on $\hat\R^{2n}$, the {\it extended system}
(\ref{r1}), (\ref{r2}) preserves the functions $\phi_1$, $\phi_2$,
the Hamiltonian \eqref{red:ham} and the reduced momentum
\begin{equation}
\label{IK} K(\gamma,p)=\langle \gamma \wedge p,\gamma \wedge
p\rangle=(\gamma,\gamma)(p,p)-(\gamma,p)^2.
\end{equation}

\subsection{Chaplygin Reducing Multiplier}
At the points of $T^*S^{n-1}$, the vector field $X=(X_\gamma,X_p)$
of the system (\ref{r1}), (\ref{r2}) can be written in the almost
Hamiltonian form $i_X(\mathbf w)=dH$, where the form $\mathbf w$
is a non-degenerate 2-form on $T^*S^{n-1}$, a semi-basic
perturbation of the canonical symplectic form
\begin{equation}
\omega=dp_1\wedge d\gamma_1+\dots+dp_n\wedge
d\gamma_n\,\vert_{T^*S^{n-1}} \label{OMEGA}\end{equation} (see
\cite{HN}).

Let $\mathbf w$ be an almost symplectic form, i.e., a
nondegenerate 2-form on an even dimensional manifold $M$. For an
almost Hamiltonian flow $\dot x=X$, $i_X\mathbf w=dH$, the {\it
Chaplygin multiplier} is a nonvanishing function $\mathcal N$ such
that $\tilde\omega=\mathcal N\mathbf w$ is closed. Since
$i_{\tilde X}\tilde\omega=dH$, $\tilde X=\frac{1}{\mathcal N}X$,
applying the time substitution $d\tau={\mathcal N}dt$, the system
$\dot x=X$ becomes the Hamiltonian system $\frac{d}{d\tau}x=\tilde
X$ with respect to the symplectic form $\tilde\omega$ \cite{St,
CCLM, Tat, EKR}. More generally, $\mathcal N$ is  the Chaplygin
multiplier if there exist a 2-form $\hat{\mathbf w}$ such that
$i_X\hat{\mathbf w}=0$ and $\tilde\omega=\mathcal N(\mathbf
w-\hat{\mathbf w})$ is symplectic (see \cite{EKR}). Then, as
above, the system $\dot x=X$ becomes the Hamiltonian system
$\frac{d}{d\tau}x=\tilde X$ with  respect to the symplectic form
$\tilde\omega$.

Alternatively, a transparent and classical way to introduce the
Chaplygin reducing multiplier for our system is as follows (e.g.,
see Section 3 in \cite{FeJo}). Let $\mathcal N(\gamma)$ be a
differentiable nonvanishing positive function in a neighborhood of
$S^{n-1}$. Consider the coordinate transformation
\begin{equation*}
(\gamma,p) \longmapsto  (\gamma,\tilde p),\qquad  \tilde
p=\mathcal{N}p
\end{equation*}
defined in some neighborhood of $T^*S^{n-1}$ and the new
symplectic form
\begin{eqnarray}\label{2-form}
\tilde\omega &=& d\tilde p_1\wedge d\gamma_1+\cdots+d\tilde
p_n\wedge d\gamma_n\vert_{T^*S^{n-1}} \\&=& \mathcal
N\omega+p_1\,d\mathcal N\wedge d\gamma_1+\dots +p_n\, d\mathcal
N\wedge d\gamma^n\,\vert_{T^*S^{n-1}}.\nonumber\end{eqnarray}

Then $\mathcal N$ is a {\it Chaplygin multiplier} for the reduced
system if the equations (\ref{r1}), (\ref{r2}) in the new time
$d\tau={\mathcal N} (q)dt$ becomes Hamiltonian with respect to the
form $\tilde \omega$. If $\mathcal N$ is a Chaplygin multiplier
then from the Liouville theorem we have
\begin{equation}\label{IM}
\mathcal L_{\tilde X}(\tilde \omega^{n-1})=0 \qquad
\Longleftrightarrow\qquad \mathcal L_X(\mathcal
N^{n-2}\omega^{n-1})=0,
\end{equation}
i.e., the original system has the invariant measure with density
$\mathcal N(\gamma)^{n-2}$. Further, the form $\mathbf w$ reads
\begin{equation}\label{nova}
\mathbf w=\omega+p_1\,d\ln\mathcal N\wedge d\gamma_1+\dots +p_n\,
d\ln\mathcal N\wedge d\gamma^n\vert_{T^*S^{n-1}}.
\end{equation}

\subsection{Homogeneous Sphere}
It is proved in \cite{HN} that the reduced equations of motion
\eqref{red:eq2} of the homogeneous ball are already Hamiltonian,
for any value of the $SO(n-1)$-momentum mapping. This interesting
result, for $\Phi=\eta=0$ can be easily derived from Theorem 2.

Suppose the inertia operator $I$ equals $s\, \mathbf I$
(multiplication by a constant $s>0$). Then the equation
(\ref{veza}), under the conditions (\ref{uslovi}), gives
$\xi={s}\,p/({s+D})$. The reduced system (\ref{r1}), (\ref{r2})
takes the form
\begin{equation}
\dot\gamma=\frac{1}{s+D}p, \qquad \dot p=-\frac{(p,p)}{s+D}\gamma,
\label{sphere}\end{equation} representing the geodesic flow of the
standard $SO(n)$-invariant metric of the sphere multiplied by
$s+D$. Note that in this case the angular velocity
$$
\omega=\frac{1}{s}(\gamma\wedge \xi)=\frac{1}{s+D}(\gamma \wedge
p)
$$
is constant along the flow of \eqref{sphere}. Actually, the
angular velocity $\omega$ is constant for the rolling of the
homogeneous ball for any value of $SO(n-1)$-momentum mapping.
Namely, substituting  $M=s\,\Omega$ into the equations \eqref{c1}
and \eqref{ch} we obtain
$$
\pr_\mathfrak h(s\dot\Omega+D\dot\Omega)=0, \qquad \pr_{\mathfrak
h^\perp}(s\dot\Omega)=0,
$$
which implies $\dot\Omega=0$. Further, from \eqref{c2},
\eqref{ch}, \eqref{oO} we get $\dot\omega=\dot{\mathbf V}=0$
 (see also
\cite{HN}).

\section{Hamiltonization}

In this section we shall perform the Hamiltonization of the
reduced Chaplygin sphere (\ref{r1}), (\ref{r2}) for the inertia
operator defined on the base $E_i \wedge E_j$ via
\begin{equation}
\label{operator} I(E_i \wedge E_j)=\frac{a_ia_jD}{D-a_ia_j} E_i
\wedge E_j, \qquad 1 \le i <j \le n,
\end{equation}
where $0< a_i a_j <D$, $1\le i,j \le n$.

The form of the inertia operator as well as the form of the
Chaplygin multiplier below is motivated by the corresponding
formulas in the problem of motion of the $n$-dimensional Veselova
problem as well as the rubber Chaplygin ball given in \cite{FeJo}
and \cite{Jo}, respectively.

Let $A=\diag(a_1,\dots,a_n)$.

In the tree-dimensional case the operator \eqref{operator} defines
a generic rigid body inertia tensor $I$. Indeed, using the
isomorphism (\ref{iso}), we get
\begin{equation}
I=\diag(I_1,I_2,I_3) \label{IOP} \end{equation} where $
{I_1}={a_2a_3D}/({D-a_2a_3}), \, {I_2}={a_3a_1D}/({D-a_3a_1}), \,
{I_3}={a_2a_3D}/({D-a_2a_3}). $

Conversely, given a generic inertia tensor (\ref{IOP}) (one can
always assume that the axes of the frame attached to the ball are
principal axes of inertia), the matrix $A=\diag(a_1,a_2,a_3)$ is
determined via
\begin{equation}\label{IOP1}
a_i=\sqrt{I_1I_2I_3D}(I_i+{D})/I_i\sqrt{(I_1+D)(I_2+D)(I_3+D)},
\qquad i=1,2,3.
\end{equation}

\begin{remark}{\rm
In general, for $n \ge 4$, the operator (\ref{operator}) is not a
physical inertia operator of a multidimensional rigid body (see
\cite{FeKo}). However, by taking conditions
\begin{equation}\label{sing}
a_1=a_2=\dots=a_{n-1}\ne a_n.
\end{equation}
 and
$2a_nD>a_1a_n+a_1D$, we get the operator $I\omega=J\omega+\omega
J$, where
\begin{equation*}
J=\diag(J_1,\dots,J_1,J_n), \quad J_1=\frac{a_1^2D}{2(D-a_1^2)},
\quad J_n=\frac{a_1a_nD}{D-a_1a_n}-\frac{a_1^2D}{2(D-a_1^2)},
\end{equation*}
representing a $SO(n-1)$-symmetric rigid body ({\it
multidimensional Lagrange case} \cite{Be}) with a {\it mass
tensor} $J$.}
\end{remark}

\begin{thm} The extended reduced Chaplygin sphere equations \eqref{r1}, \eqref{r2}, defined by the inertia tensor
\eqref{operator}, read
\begin{eqnarray}
&& \dot
\gamma=\frac{1}{D}p-\frac{(p,\gamma)}{D(\gamma,A^{-1}\gamma)}A^{-1}\gamma+\frac{(\gamma,Ap)}{D^2(\gamma,A^{-1}\gamma)}\gamma
-\frac{(\gamma,\gamma)}{D^2(\gamma,A^{-1}\gamma)}Ap,
\label{ro1}\\
&&\dot
p=\frac{(p,A^{-1}\gamma)}{D(\gamma,A^{-1}\gamma)}p-\frac{(p,p)}{D(\gamma,A^{-1}\gamma)}A^{-1}\gamma+\frac{(p,Ap)}{D^2(\gamma,A^{-1}\gamma)}\gamma
-\frac{(p,\gamma)}{D^2(\gamma,A^{-1}\gamma)}Ap.\label{ro2}
\end{eqnarray}
\end{thm}

\begin{proof} From the definition (\ref{operator}), the angular velocity
is given by
\begin{equation}
\label{operator*}
\omega=I^{-1}(\gamma\wedge\xi)=A^{-1}\gamma\wedge
A^{-1}\xi-\frac{1}{D}\gamma\wedge \xi.
\end{equation}

Now, the equation (\ref{veza}), under the conditions
(\ref{uslovi}), can be solved
\begin{equation}\label{xi}
\xi=\frac1{D(\gamma,A^{-1}\gamma)} \left(Ap - (p,A\gamma)
\gamma\right)\, .
\end{equation}
Thus $\omega= \left(A^{-1}\gamma \wedge p - \gamma\wedge
Ap/D\right)/D(\gamma,A^{-1}\gamma)$ and (\ref{ro1}), (\ref{ro2})
simply follows from (\ref{r1}), (\ref{r2}). \end{proof}

\subsection{Invariant Measure} The canonical volume form $\varOmega$ on $\R^{2n}$ induces
the volume form $\sigma$ on $T^*S^{n-1}\subset \R^{2n}$ (e.g., see
paragraph 3.6, Ch. 1 \cite{AKN}). A simple description of
$\sigma$, in terms of the restricted symplectic structure
(\ref{OMEGA}) is as follows.

 Consider the standard spherical coordinates
$(\theta,r)=(\theta_1,\dots,\theta_{n-1},r)$ on $\R^n(\gamma)$ and
the corresponding canonical momenta
$(\pi_\theta,\pi_r)=(\pi_1,\dots,\pi_{n-1},\pi_r)$ on
$\R^{2n}(\gamma,p)$ with respect to the canonical symplectic form:
$$
dp_1\wedge d\gamma_1+\cdots+dp_n\wedge d\gamma_n=d\pi_1\wedge
d\theta_1+\cdots+d\pi_{n-1}\wedge d\theta_{n-1}+d\pi_r\wedge dr.
$$

Then the volume form $\varOmega$ can be represented as
\begin{equation}
\varOmega= dp_1\wedge d\gamma_1\wedge \cdots \wedge dp_n\wedge
d\gamma_n= (d\pi_1\wedge d\theta_1\wedge\dots\wedge
d\pi_{n-1}\wedge d\theta_{n-1})\wedge dp_r \wedge dr,
\label{VOLUME}
\end{equation}
where $r=\sqrt{(\gamma,\gamma)}$ and
$p_r=(\gamma,p)/\sqrt{(\gamma,\gamma)}$. The coordinates
$(\theta,\pi_\theta)$ are canonical coordinates (the symplectic
form (\ref{OMEGA}) equals $d\pi_1\wedge
d\theta_1+\cdots+d\pi_{n-1}\wedge d\theta_{n-1}$) and
$$
\sigma=d\pi_1\wedge d\theta_1\wedge\dots\wedge d\pi_{n-1}\wedge
d\theta_{n-1}
$$
is the canonical volume form on the cotangent bundle $T^*S^{n-1}$,
naturally extended to $\hat\R^{2n}$.

\begin{prop}
\label{redLR} The reduced Chaplygin system \eqref{ro1},
\eqref{ro2} on $T^*S^{n-1}$ possesses an invariant measure
\begin{equation}
\mu(\gamma)\,\sigma\,=(A^{-1}\gamma,\gamma)^{-(n-2)/2}\, \sigma\,.
\label{DENSITY}\end{equation}
\end{prop}

\begin{proof} The divergence of the vector field $X$ in $\hat\R^{2n}$ is
\begin{equation}
\mathrm{div}(X)=\sum_{i=1}^n \left( \frac{\partial \dot
\gamma_i}{\partial \gamma_i} +\frac{\partial \dot p_i}{\partial
p_i}\right) =(n-2) \left(\frac{(\gamma,
A^{-1}p)}{D(\gamma,A^{-1}\gamma)}+\frac{(\gamma,Ap)}{D^2(\gamma,A^{-1}\gamma)}\right)+\Psi,
\label{div1}
\end{equation}
where
$$
\Psi=\left(\frac{2(A^{-2}\gamma,\gamma)}{D(\gamma,A^{-1}\gamma)^2}+\frac{2(\gamma,\gamma)}{D^2(\gamma,A^{-1}\gamma)^2}
-\frac{\mathrm{tr}
A^{-1}}{D(\gamma,A^{-1}\gamma)}-\frac{\mathrm{tr}
A}{D^2(\gamma,A^{-1}\gamma)})\right)(\gamma,p).
$$

Whence, on the invariant submanifold $\phi_2=\pi_r=0$, in view of
(\ref{ro1}), we get
$$
\sum_{i=1}^n \left( \frac{\partial \dot \gamma_i}{\partial
\gamma_i} +\frac{\partial \dot p_i}{\partial p_i}\right)=(n-2)
\frac{(\gamma, A^{-1}\dot
\gamma)}{(\gamma,A^{-1}\gamma)}=-\frac{\dot\mu}{\mu}.
$$
In other words, the density $\mu(\gamma)$ satisfies the Liouville
equation
\begin{equation}
\mathrm{div}(\mu X)=\sum_{i=1}^n \dot q_i\frac{\partial
\mu}{\partial q_i}+\mu\sum_{i=1}^n \left( \frac{\partial \dot
q_i}{\partial q_i} +\frac{\partial \dot p_i}{\partial
p_i}\right)=0 \label{div2}
\end{equation}
on the manifold $\phi_2=\pi_r=0$.

On the other side,  from (\ref{VOLUME}) we obtain
\begin{equation}
\mathcal L_X(\mu\varOmega)=\mathcal L_X(\mu\sigma)\wedge d\pi_r
\wedge dr+ \mu\sigma \wedge \mathcal L_X(d\pi_r\wedge dr).
\label{div3}
\end{equation}

Since the functions $\phi_1$, $\phi_2$ are invariants of the
vector field $X$, the Lie derivatives ${\mathcal L}_X d\pi_r$ and
${\mathcal L}_X dr$  equal zero. Further, (\ref{div2}) implies
that the left hand side of (\ref{div3}) is also equal to zero on
the invariant subvariety $\phi_2=\pi_r=0$. Thus we conclude
$$
\mathcal L_X(\mu\sigma)\vert_{T^*S^{n-1}}=0
$$ as required.  \end{proof}

\begin{remark}{\rm
The reduced vector field \eqref{red:eq2} has an invariant measure
for $\Phi=\eta\ne 0$ as well. Namely, the $SO(n-1)$-reduced system
 \eqref{PE}, \eqref{ch-red} preserve the volume form \eqref{mu*}. Then the restriction of the flow to the invariant manifold
$M_\eta$ (see diagram \eqref{sym:red}) preserves the {\it induced
volume form} (e.g., see \cite{AKN}). In this sense, Proposition
\ref{redLR} is equivalent to the proportionality of the densities
of measures \eqref{mu*} and \eqref{DENSITY} (compare with Theorem
5.1 \cite{FeJo}). In particular, for $n=3$ the density
$\mu(\gamma)$ after the transformation \eqref{IOP1}, up to a
multiplication by a constant, takes the form
\eqref{mu-ch}.}\end{remark}

\subsection{Time Reparametrization}
The reduced Hamiltonian \eqref{red:ham} read
\begin{equation}
H(\gamma,p)=\frac{1}{2D(\gamma,A^{-1}\gamma)} \langle \gamma
\wedge p, A^{-1}\gamma \wedge p-\frac{1}{D}\gamma \wedge Ap
\rangle.\label{ham0}
\end{equation}
According to the constraints (\ref{phi}), instead of (\ref{ham0})
we can use the Hamiltonian function
\begin{equation}
H(\gamma,p)=\frac{1}{2D^2(\gamma,A^{-1}\gamma)}\left(D(\gamma,A^{-1}
\gamma)(p,p)-(p,Ap)\right)\label{hamiltonian}
\end{equation}

As follows from Proposition \ref{redLR} and the relation
(\ref{IM}), if the reduced Chaplygin system on $T^* S^{n-1}$ is
transformable to a Hamiltonian form by a time reparameterization,
then the corresponding reducing multiplier $\N$ should be
proportional to $1/\sqrt{(\gamma, A^{-1}\gamma)}$.

\begin{thm}
\label{main} Under the time substitution
\begin{equation}\label{CM}
d\tau =\mathcal N \,dt=\frac1{D\sqrt{(A^{-1}\gamma,\gamma) }}\,\,
dt
\end{equation}
and an appropriate change of momenta
\begin{equation}
(\gamma,p) \longmapsto (\gamma,\tilde p) , \qquad \tilde
p=\frac1{D\sqrt{(\gamma, A^{-1}\gamma)}}\,p \label{mapping}
\end{equation}
the reduced system \eqref{ro1}, \eqref{ro2} becomes a Hamiltonian
system describing a geodesic flow on $T^*S^{n-1}$ with the
Hamiltonian
\begin{equation}
H(\gamma,\tilde p)=\frac{1}{2}\left(D(\gamma,A^{-1} \gamma)(\tilde
p,\tilde p)-(\tilde p,A\tilde p)\right).\label{hamiltonian*}
\end{equation}
\end{thm}

\begin{proof} Consider the cotangent bundle $T^*S^{n-1}$ realized as a
submanifold of $\R^{2n}$ given by
\begin{equation}
\psi_1\equiv (\gamma,\gamma)=1, \quad \psi_2\equiv (\gamma,\tilde
p)=0. \label{psi}
\end{equation}

The canonical Poisson bracket on $T^*S^{n-1}$ with respect to the
symplectic form (\ref{2-form}) can be described by the use of the
Dirac bracket (see \cite{Dirac, Mo, AKN}):
$$
\{F, G\}_d =\{F, G\} -(\{F,\psi_1 \}\{G,\psi_2\}- \{F,\psi_2\}
\{G,\psi_1\} )/ { \{\psi_1, \psi_2 \} },
$$
where $$ \{F,G\}=\sum_{i=1}^n\left(\frac{\partial
F}{\partial\gamma_i} \frac{\partial G}{\partial\tilde
p_i}-\frac{\partial F}{\partial\tilde p_i}\frac{\partial
G}{\partial\gamma_i}\right). $$

Considered on $\hat\R^{2n}$, the bracket $\{\cdot,\cdot\}_d$ is
degenerate and has two Casimir functions $\psi_1$ and $\psi_2$.
The symplectic leaf given by (\ref{psi}) is exactly the cotangent
bundle $T^*S^{n-1}$ endowed with the canonical symplectic form.

Under the mapping (\ref{mapping}), the Hamiltonian
(\ref{hamiltonian}) transforms to (\ref{hamiltonian*}). With the
above notation, the geodesic flow defined by Hamiltonian function
(\ref{hamiltonian*}), in the time $\tau$, is the restriction to
(\ref{psi}) of
\begin{equation}
\gamma_i^\prime=\frac{d}{d\tau} \gamma_i=\{\gamma_i, H\}_d, \quad
\tilde p_i^\prime=\frac{d}{d\tau}\tilde p_i=\{\tilde p_i, H\}_d ,
\quad i=1,\dots,n. \label{hamiltonian-flow}
\end{equation}

It is convenient to find equations (\ref{hamiltonian-flow}) using
the Lagrange multipliers (see \cite{Mo, AKN}). Introduce
$$
H^*=H-\lambda \psi_1-\mu \psi_2.
$$
The equations (\ref{hamiltonian-flow}) are then given by
\begin{eqnarray*}
&&\gamma^\prime=\frac{\partial H^*}{\partial \tilde p}=
\frac{\partial H}{\partial \tilde p}-\mu \gamma=D(A^{-1}\gamma,\gamma)\tilde p-A\tilde p -\mu \gamma,\\
&&\tilde p^\prime=-\frac{\partial H^*}{\partial
\gamma}=-\frac{\partial H}{\partial \gamma}+2\lambda
\gamma+\mu\tilde p= -D(\tilde p,\tilde
p)A^{-1}\gamma+2\lambda\gamma+\mu \tilde p
\end{eqnarray*}
where the multipliers $\lambda$ and $\mu$ are determined from the
condition that the constraint functions $\psi_1$ and $\psi_2$ are
integrals of the motion.

Straightforward calculations yield
$$
\lambda=\frac{(A\tilde p,\tilde p)}{2(\gamma,\gamma)}, \qquad
\mu=\frac{D(\gamma,A^{-1}\gamma)(\tilde p,\gamma)-(A\tilde
p,\gamma)}{(\gamma,\gamma)}
$$
and therefore
\begin{eqnarray}
&& \gamma^\prime=D(A^{-1}\gamma,\gamma)\tilde p- A\tilde p+
\frac{(\gamma,A\tilde p)}{(\gamma,\gamma)} \gamma-\frac{D(\gamma,A^{-1}\gamma)(\tilde p,\gamma)}{(\gamma,\gamma)}\gamma, \label{ham_q}\\
&& \tilde p^\prime= -D(\tilde p,\tilde
p)A^{-1}\gamma+\frac{(\tilde p,A\tilde p)}{(\gamma,\gamma)}
\gamma-\frac{(\gamma,A\tilde p)}{(\gamma,\gamma)} \tilde
p+\frac{D(\gamma,A^{-1}\gamma)(\tilde
p,\gamma)}{(\gamma,\gamma)}\tilde p. \label{ham_p}
\end{eqnarray}

In the time $t$,  inverting the mapping (\ref{mapping}), the
equation (\ref{ham_q}) takes the form
\begin{eqnarray*}
\dot \gamma\cdot
D\sqrt{(\gamma,A^{-1}\gamma)} &=& \frac1{D\sqrt{(\gamma,A^{-1}\gamma)}}\\
&& \cdot \left( D(A^{-1}\gamma,\gamma)p- A p+
\frac{(\gamma,Ap)}{(\gamma,\gamma)} \gamma
-\frac{D(\gamma,A^{-1}\gamma)(p,\gamma)}{(\gamma,\gamma)}\gamma\right),
\end{eqnarray*}
i.e.,
\begin{equation}
\dot \gamma=\frac{1}{D}p- \frac{1}{D^2(\gamma,A^{-1}\gamma)}Ap+
\frac{(\gamma,Ap)}{D^2(\gamma,A^{-1}\gamma)(\gamma,\gamma)}
\gamma-\frac{(p,\gamma)}{D(\gamma,\gamma)}\gamma, \label{*}
\end{equation} which
coincides with (\ref{ro1}) at the points of $T^*S^{n-1}$. Further,
\begin{eqnarray}
\frac{d}{d\tau}\tilde p &=&
\frac{d}{d\tau}\left(\frac{p}{D\sqrt{(\gamma,A^{-1}\gamma)}}\right)=
\frac{d}{dt}\left(\frac{p}{D\sqrt{(\gamma,A^{-1}\gamma)}}\right)D\sqrt{(\gamma,A^{-1}\gamma)}
 \nonumber\\
&=& \left(
p\frac{d}{dt}\frac{1}{\sqrt{(\gamma,A^{-1}\gamma)}}+\dot
p\frac{1}{\sqrt{(\gamma,A^{-1}\gamma)}}\right)\sqrt{(\gamma,A^{-1}\gamma)}=
\dot p-p\frac{(A^{-1}\gamma,\dot \gamma)}{(\gamma,A^{-1}\gamma)}.
\label{***}\end{eqnarray}

Finally, substituting $\tilde p=\N p$ into the right hand side of
(\ref{ham_p}), combining with (\ref{*}) and (\ref{***}), we get
\begin{eqnarray}\label{**}
\dot p  &=&  -\frac{(p,p)}{D(\gamma,A^{-1}\gamma)}
A^{-1}\gamma+\frac{(p,A
p)}{D^2(\gamma,A^{-1}\gamma)(\gamma,\gamma)} \gamma \\\nonumber
&&+\frac{(p,A^{-1}\gamma)}{D(\gamma,A^{-1}\gamma)}p-
\frac{(\gamma,p)}{D^2(\gamma,A^{-1}\gamma)^2}p.\end{eqnarray}

As above,  the equations (\ref{ro2}) and (\ref{**}) are different,
but they coincide on the invariant manifold $\phi_1=\psi_1=1$,
$\phi_2=\psi_2=0$. The theorem is proved. \end{proof}

\begin{remark}\label{referee}{\rm
The link between the Dirac bracket and the Lagrange multiplier
approach can be expressed via
$$
\lambda=\frac{\{H,\psi_2\}}{\{\psi_1,\psi_2\}}, \qquad
\mu=-\frac{\{H,\psi_1\}}{\{\psi_1,\psi_2\}}.
$$
Also, note that the reduced almost symplectic form \eqref{nova} is
given by:
$$
\mathbf w=\sum_{i,j=1}^n dp_i \wedge
d\gamma_i-\frac{p_ia_j^{-1}\gamma_j}{(\gamma,A^{-1}\gamma)}
d\gamma_j\wedge d\gamma_i \ \vert_{T^*S^{n-1}}.
$$
}\end{remark}

\begin{remark}{\rm During the referee process of this paper, the paper \cite{H}
appeared, where the Abelian $\mathfrak v$-Chaplygin systems
associated to Cartan decompositions $\mathfrak g=\mathfrak k
\oplus \mathfrak p$ of semi-simple Lie algebras are studied. They
are defined on the direct product of a Lie group $K$ ($\mathfrak
k=Lie(K)$) endowed with a left-invariant metric with the vector
space $\mathfrak v$ ($\mathfrak v=[\Gamma,\mathfrak k]\subset
\mathfrak p$) endowed with the metric induced from the Killing
form. Here $\Gamma\in\mathfrak p$ is fixed. As an example, taking
the Cartan decomposition $so(n,1)=so(n)\oplus \R^n$ of the Lie
algebra $so(n,1)$ one gets the Chaplygin sphere problem (compare
with the equations (73) in \cite{Jo}). Besides $\mathfrak
v$-reduction to $T^*K$, likewise the Chaplygin sphere problem, the
system has an internal symmetry group $H\subset K$ (isotropy group
of $\Gamma$) and admits the almost symplectic reduction with
respect to the $H$-action. Hochgerner derived the equations on the
parameters of the kinetic energy, such that the (zero momentum)
reduced almost symplectic form is conformally symplectic. The
operator \eqref{operator} represents the solution of these
equations within the class of diagonal operators on $so(n)$ with
respect to the base $E_i\wedge E_j$ \cite{H}.}
\end{remark}

\section{Integrability}

\subsection{Classical Chaplygin Sphere and the Veselova Problem}

The \textit{Veselova problem} describes the motion of a rigid body
about a fixed point subject to  the nonholonomic constraint
\begin{equation}
\label{classical} (\vec{\mathbf w},\vec\gamma )=0,
\end{equation}
where $\vec{\mathbf w}$ is the vector of the angular velocity in
the body frame and $\vec\gamma$ is a representation of a unit
vector fixed in a space, relative to the body frame \cite{VeVe2}.
The equations of motion in the moving frame have the form
\begin{equation}
\frac{d}{dt}{\mathcal I}\vec{\mathbf w}={\mathcal I}\vec{\mathbf
w} \times \vec{\mathbf w} + \lambda \vec\gamma,  \qquad
\frac{d}{dt}\vec\gamma= \vec\gamma \times \vec{\mathbf w},
\label{3.1}
\end{equation}
where $\mathcal I$ is the inertia tensor of the rigid body and
$\lambda$ is a Lagrange multiplier chosen such that $\vec{\mathbf
w}(t)$ satisfies the constraint (\ref{classical}),
\begin{equation} \label{la}
\lambda= - \frac {({\mathcal I}\vec{\mathbf w} \times \vec{\mathbf
w}, {\mathcal I}^{-1}\vec\gamma)}{({\mathcal
I}^{-1}\vec\gamma,\vec\gamma ) }\, .
\end{equation}
Here we suppose that all eigenvalues of $\mathcal I$ are greater
then $1$.

Equations (\ref{3.1}), (\ref{la}) also define a dynamical system
on the whole space $\R^6(\vec{\mathbf w},\vec\gamma)$, and the
constraint function $ f_1=(\vec{\mathbf w},\vec\gamma)$ appears as
its first integral. The system has an invariant measure with
density $\sqrt{ ({\mathcal I}^{-1}\vec\gamma,\vec\gamma )}$.
Following \cite{Fe1}, by introducing $ \vec K=\mathcal
I\vec{\mathbf w}-(\mathcal I^0\vec{\mathbf
w},\vec\gamma)\vec\gamma$, $\mathcal I^0=\mathcal I-\mathbf I $
one can write system (\ref{3.1}), (\ref{la}) in the form
\begin{equation}
\frac{d}{dt}\vec K=\vec K \times \vec{\mathbf w}, \qquad
\frac{d}{dt}\vec\gamma= \vec\gamma \times \vec{\mathbf w}.
\label{3.2}
\end{equation}

Apart from $f_1=(\vec{\mathbf w},\vec\gamma)=(\vec K,\vec\gamma)$,
it always has the geometric integral
$f_2=(\vec\gamma,\vec\gamma)=1$ and two other independent
integrals
\begin{equation}\label{ints}
f_3=\frac12(\vec K,\vec{\mathbf w})-\frac12(\vec
K,\vec\gamma)(\mathcal I^0\vec{\mathbf w},\vec\gamma), \qquad
f_4=(\vec K,\vec K).\end{equation}

On the constraint subvariety (\ref{classical}), these functions
reduce to the energy integral $\frac12 ({\mathcal I}\vec{\mathbf
w},\vec{\mathbf w})$ and $(I\vec{\mathbf w}, I\vec{\mathbf w})-
(I\vec{\mathbf w},\vec\gamma)^2$ (see \cite{VeVe2}).

By the {Euler-Jacobi theorem} \cite{AKN}, the above system is
solvable by quadratures on the whole space ${\mathbb R}^6$. For
$f_1=0$ the system was integrated by Veselova (e.g., one can find
the motion using the isomorphism with a celebrated Neumann system
\cite{VeVe2}). Next, as was shown in \cite{Fe1}, the restriction
of the extended Veselova system \eqref{3.1}, \eqref{la} onto the
level variety $f_1=c_1$ ($c_1\ne 0$) can be reduced to this system
on the level $f_1=0$ by a linear change of variables $(\vec
K,\vec\gamma)\longmapsto (\vec K_1,\vec\gamma_1)$ and an
appropriate time reparametization. This linear change was found by
using a relation of the Veselova system with the Chaplygin sphere
problem, which we are going to describe now.

Define the operator $I$ and vector $\vec{\omega}$ by:
\begin{equation}\label{tr1}
\mathcal I=\mathbf I+DI^{-1}, \quad \vec{\mathbf w}=-I\vec{\omega}
\quad \Longleftrightarrow \quad I=D(\mathcal I-\mathbf I)^{-1},
\quad \vec{\omega}= - \frac{1}{D}(\mathcal I-\mathbf
I)\vec{\mathbf w}.
\end{equation}

Now we can state the following remarkable correspondence:

\begin{thm} \label{fedorov} {\rm (Fedorov\, \cite{Fe1})}
The invariant tori $ f_1=c_1$, $f_2=1$, $ f_3=c_3$, $f_4=c_4 $ of
the Veselova problem \eqref{3.1}, \eqref{la},  via \eqref{tr1}
transform to the invariant tori $ F_1=c_1$, $ F_2=1$, $ F_3=c_3$,
$ F_4=c_4 $ of the Chaplygin sphere problem \eqref{Chap}.
\end{thm}

Let us mention that there are two interesting isomorphisms between
the Chaplygin sphere problem \eqref{Chap} with $F_1=(\vec{\mathbf
k},\vec\gamma)=0$ and the Clebsh case of the Kirchoffs equations
of a rigid body motion in an ideal fluid, with a zero area
integral. The first one is described in \cite{Fed:dis} and the
other one is given recently in \cite{BM3}.

\subsection{Veselova Problem on $SO(n)$}
It appears that the analogue of Theorem \ref{fedorov} can be
formulated for an arbitrary dimension $n$ and a zero value of the
momentum \eqref{zero}. First, for a reader's sake, we shall
briefly recall some definitions and results of \cite{FeJo}.

Consider a nonholonomic LR system on $SO(n)$ defined by the
left-invariant Lagrangian $ L_\mathcal I(g,\dot g)=\frac12\langle
{\mathcal I}\mathbf w,\mathbf w\rangle=
-\frac14\,\mbox{tr}(\mathcal I\mathbf w\mathbf w)$ where
${\mathcal I}\,:\, so(n)\to so(n)$ is positive definite and the
right-invariant distribution $\mathcal D_r$ on $TSO(n)$ whose
restriction to the algebra $so(n)$ is given by $
\mathfrak{d}=\Span\{E_i \wedge E_j \; \vert\,
i=1,\dots,r,\,j=1,\dots,n \}. $ This implies the constraints
\begin{equation}
\langle \mathbf w,\, \Ad_{g^{-1}}(E_i \wedge E_j)\rangle=\langle
\mathbf w, \,e_i\wedge e_j\rangle=0, \qquad n-r+1\le i<j\le n.
\label{VC}
\end{equation}

Here $\mathbf w(t)=g^{-1}\cdot g(t) \in so(n)$ and $
e_1=(e_{11},\dots,e_{1n})^T,\dots,e_n=(e_{n1},\dots,e_{nn})^T $
is the  orthogonal frame of unit vectors fixed in the space {\it
and regarded in the moving frame} ($ E_1=g\cdot e_1,\; \dots,\;
E_n=g\cdot e_n, $ where $E_1=(1,0,\dots,0)^T,\; \dots,\;
E_n=(0,\dots,0,1)^T$). They play the role of redundant coordinates
on $SO(n)$.

The system is described by the kinematic Poisson equations
\begin{equation}
\label{PoissonV} \dot e_i= - \mathbf w e_i, \qquad  i=1,\dots,n,
\end{equation}
together with the Euler-Poincar\'e  equations with indefinite
multipliers $\lambda_{pq}$
\begin{equation}
\frac{d}{dt}\left({\mathcal I}\mathbf w\right) =[{\mathcal
I}\mathbf w,\mathbf w] +\sum_{n-r+1\le p<q\le n}\lambda_{pq} \,
e_p\wedge e_q. \label{VFK}
\end{equation}

Since for $n=3$, $r=2$ the above system represents
\textit{Veselova problem}, we refer to $(SO(n),L_\mathcal
I,\mathcal D_r)$ as a {\it generalized Veselova system} (see
Fedorov and Kozlov \cite{FeKo}).

The Lagrangian $L_\mathcal I$ and the distribution  $\mathcal D_r$
are invariant with respect to the left $SO(n-r)$-action, where
$SO(n-r)$ is the subgroup of $SO(n)$, rotations that leave
$E_1,\dots,E_r$ invariant. Moreover, the distribution $\mathcal
D_r$ can be seen as a principal connection of the bundle
\begin{equation*}\begin{array}{ccc}
SO(n-r) & \longrightarrow & SO(n)\qquad\qquad\qquad\qquad\,\,  \\
&  & \downarrow  \qquad\qquad\qquad\qquad\,\,\,\, \\
&  &  V_{n,r}=SO(n)/SO(n-r).
\end{array}
\label{bundle}
\end{equation*}
As a result, the  system can naturally be regarded as a Chaplygin
system and dynamics is reducible to the Stiefel variety $V_{n,r}$.
The points  of the Stiefel variety can be seen as matrices
$\mathcal X=(e_1,\dots,e_r)$ (positions of the $r$-frame given by
vectors $(e_1,\dots,e_r)$). So, the tangent bundle $TV(r,n)$ is
the set of pairs $({\mathcal X}, \dot {\mathcal X})$ of $n\times
r$ matrices subject to the constraints
\begin{equation} \label{cond_X}
{\mathcal X}^T {\mathcal X}={\bf I}_r, \quad {\mathcal X}^T
\dot{\mathcal X} + \dot{\mathcal X}^T {\mathcal X} =0.
\end{equation}

The reduced Lagrangian takes the form $L_{red}({\mathcal X},
\dot{\mathcal X}) =- \frac 1{4}\mathrm{tr}\left({\mathcal I}
\Phi_r\Phi_r\right)$ (see \cite{FeJo}), where the tangent bundle
momentum mapping $\Phi_r\, :\, TV_{n,r} \to so(n)\cong so(n)^*$ is
given by
\begin{eqnarray*}
\Phi_r( {\mathcal X} ,\dot {\mathcal X} ) &=&  {\mathcal X} \dot
{\mathcal X}^T-  \dot {\mathcal X}  {\mathcal X}^T +\frac 12
{\mathcal X} [ {\mathcal X}^T \dot  {\mathcal X}
-  \dot {\mathcal X}^T {\mathcal X}] {\mathcal X}^T  \nonumber \\
&=& e_1 \wedge \dot e_1+\cdots +  e_r \wedge \dot e_r +\frac 12
\sum_{1\le\alpha<\beta\le r} \left[  (e_\alpha, \dot e_\beta)-
(\dot e_\alpha, e_\beta)\right] \, e_\alpha \wedge e_\beta .
\end{eqnarray*}

Introduce the $n \times r$ momentum matrix
\begin{equation}
{\mathcal P}_{is}=\partial L_{red}({\mathcal X}, \dot{\mathcal X})
/\partial \dot{\mathcal X}_{is}\, . \label{dotX->P}
\end{equation}
Since the Lagrangian is degenerate in the redundant velocities
$\dot{\mathcal X}_{is}$, from this relation one cannot express
$\dot{\mathcal X}$ in terms of $({\mathcal X}, {\mathcal P})$
uniquely. On the other hand, the cotangent bundle $T^*V(r,n)$ can
be realized as the set of pairs $({\mathcal X}, {\mathcal P})$
satisfying the constraints
\begin{equation} \label{condX}
{\mathcal X}^T {\mathcal X} ={\bf I}_r, \quad {\mathcal X}^T
{\mathcal P} + {\mathcal P}^T {\mathcal X} =0\, .
\end{equation}
Under the conditions \eqref{cond_X}, \eqref{condX}, the relation
(\ref{dotX->P}) can be uniquely inverted, i.e., one gets
$\dot{\mathcal X}=\dot{\mathcal X}(\mathcal X,\mathcal P)$. Then
we have (see Theorem 5.4 in \cite{FeJo}):

\begin{thm} {\rm (\cite{FeJo})}\label{rest_St}
The $SO(n-r)$-reduction of the Veselova problem \eqref{VC},
\eqref{PoissonV}, \eqref{VFK} is the restriction to $T^*V(r,n)$ of
the following system on the space $({\mathcal X}, {\mathcal P})$:
\begin{equation} \label{dot_P}
\dot{\mathcal X}=-\Phi_r ({\mathcal X}, {\mathcal P})\, {\mathcal
X}, \quad \dot{\mathcal P}= -\Phi_r ({\mathcal X}, {\mathcal P})\,
{\mathcal P},
\end{equation}
where $\Phi_r({\mathcal X}, {\mathcal P})=\Phi_r({\mathcal X},
\dot{\mathcal X}({\mathcal X}, {\mathcal P}))$.
\end{thm}

\begin{remark}{\rm
Here we use the opportunity to mention one correction to
\cite{FeJo}: in the equation (5.21) the momentum mapping $\Phi^*$
should read $\Phi^*=\mathcal I{\omega}\vert_{\mathcal
D_r}=\mathcal X\mathcal P^T-\mathcal P\mathcal X^T$. This equation
was used only in the proof of Theorem 5.4 \cite{FeJo}. The
statement of the theorem itself remains to be correct.
}\end{remark}

In particular, for  $r=1$, the Veselova problem is reducible to
$T^*S^{n-1}$.

Let, as above, $A=\diag(a_1,\dots,a_n)$ and denote
$\gamma=\mathcal X=e_1$, $p=\mathcal P$. Taking the special
inertia operator defined by
\begin{equation}
\mathcal{I} (E_i\wedge E_j)={a_i^{-1} a_j^{-1}}\,E_i\wedge E_j,
\qquad 1 \le i <j \le n, \label{inertia_tensor}
\end{equation}
we have $ \Phi_1=\gamma\wedge \dot \gamma=\gamma\wedge Ap
/{(\gamma,A^{-1}\gamma)}$ and the reduced system \eqref{dot_P}
becomes (here we replaced the matrix $A$ from \cite{FeJo, FeJo3}
by $A^{-1}$)
\begin{eqnarray}
&&\dot \gamma =-\Phi_1\,\gamma=\frac1{(\gamma,A^{-1}\gamma)}
\left(- {(p,A\gamma)} \gamma + (\gamma,\gamma) Ap
\right) ,  \label{extended1} \\
&&\dot p =-\Phi_1 \, p =  \frac1{(\gamma,A^{-1}\gamma)}
\left((p,Ap)\gamma - (p,\gamma)A p \right)\label{extended2}.
\end{eqnarray}

Furthermore, as it follows from \cite{FeJo, FeJo3}, under the time
substitution (\ref{CM}) and the change of momenta (\ref{mapping})
the reduced system  transforms to a Hamiltonian system describing
an integrable geodesic flow on $T^*S^{n-1}$ with the Hamiltonian
\begin{equation}
\mathcal H(\gamma,\tilde p)=\frac{D^2}{2}(A\tilde p,\tilde p).
\label{ham:ves}
\end{equation}

\begin{remark} \label{elipsoid}{\rm The reduced Veselova system \eqref{extended1}, \eqref{extended2} is
trajectory equivalent to the geodesic flow on the ellipsoid $
E^{n-1}=\{x=(x_1,\dots,x_n)\in {\mathbb R^n}\mid (x,Ax)=1\} $: the
geodesic lines $x(t)$ of the ellipsoid, after the Gauss mapping
$\gamma(t)={A x(t)}/\vert A\gamma(t)\vert$ and a time rescaling,
become solutions of the reduced Veselova system, and vice versa
(see \cite{FeJo3}). }\end{remark}

\subsection{Integrability of the Reduced Chaplygin Sphere Problem}
Let us suppose $a_i \ne a_j$, $i\ne j$. As in the
three-dimensional case \cite{Fe1}, we have

\begin{thm}\label{CH-VP}
{\rm (i)} The geodesic flow \eqref{ham_q}, \eqref{ham_p} is
completely integrable on $T^*S^{n-1}$.

{\rm (ii)} The zero $SO(n-1)$-momentum reduced multidimensional
nonholonomic Chaplygin sphere problem \eqref{ro1}, \eqref{ro2}
defined by inertia operator \eqref{operator} and the reduced
Veselova problem \eqref{extended1}, \eqref{extended2} defined by
inertia operator \eqref{inertia_tensor} have the same invariant
toric foliation of $T^*S^{n-1}$.

{\rm (iii)} Let $\mathbf T$ be a regular, $(n-1)$-dimensional
invariant torus. Then there exist angle coordinates
$\varphi_1,\dots,\varphi_{n-1}$ on $\mathbf T$ in which both
problems simultaneously take the form
\begin{equation*}
\dot \varphi_1= \frac{\omega_1^i}{D\sqrt{(A^{-1}\gamma,\gamma)
}},\,\, \dots\, ,
\dot\varphi_{n-1}=\frac{\omega_{n-1}^i}{D\sqrt{(A^{-1}\gamma,\gamma)
}}\,  \label{linearizacija}
\end{equation*} with frequencies
$\omega^i_1,\dots,\omega_{n-1}^i$, $i=1,2$, respectively.
\end{thm}

\begin{proof} In what follows, we restrict our considerations to
$T^*S^{n-1}$. The momentum integral (\ref{IK}) in variables
$(\gamma,\tilde p)$ becomes
\begin{equation}
K(\gamma,\tilde p)=D^2(A^{-1}\gamma,\gamma)(\tilde p,\tilde
p)\label{kk}
\end{equation}
and the Hamiltonian \eqref{hamiltonian*} can be written in the
form
$$
H(\gamma,\tilde p)=\frac{1}{2D}K(\gamma,\tilde
p)-\frac{1}{D^2}\mathcal H(\gamma,\tilde p).
$$

Since \eqref{kk} is the integral of the geodesic flow
(\ref{ham_q}), (\ref{ham_p}), on $T^*S^{n-1}$ we have
\begin{equation*}\label{komutiranje}
\{H,K\}_d=\{H,\mathcal H\}_d=\{\mathcal H, K\}_d=0.
\end{equation*}

Consider the spheroconical coordinates
$(\lambda_1,\dots,\lambda_{n-1})$ ($a_1<\lambda_{1}<a_2< \dots
<\lambda _{n-1}<a_n$) on $S^{n-1}$ defined by the relations
\begin{equation} \label{S_n}
\gamma_{i}^{2}=\frac{\left( a_{i}-\lambda _{1}\right) \cdots
\left( a_{i}-\lambda_{n-1}\right) }{\prod_{j\ne i} \left( a_{i}
-a_{j}\right) }, \qquad i=1,\dots,n
\end{equation}
(see \cite{Mo}). Let $(\mu_1,\dots,\mu_{n-1})$ be the canonical
momenta on the cotangent bundle with respect to the form
(\ref{2-form})
$$
\tilde\omega = d\tilde p_1\wedge d\gamma_1+\cdots+d\tilde
p_n\wedge d\gamma_n\vert_{T^*S^{n-1}}=d\mu_1\wedge
d\lambda_1+\dots+d\mu_{n-1}\wedge d\lambda_{n-1}.
$$
Then, according to \cite{Mo}, \cite{Br} and \cite{FeJo},
respectively, we have:
\begin{eqnarray*}
&&(\tilde p,\tilde p)= -4\sum\limits_{k=1}^{n-1} \frac{\left(
\lambda_{k}-a_{1}\right) \cdots \left( \lambda _{k}-a_{n}\right)}
{\prod_{s\neq k}\left( \lambda _{k}-\lambda_{s}\right) }\mu_{k}^{2}, \\
&&(\gamma,A^{-1}\gamma)=  \frac{\lambda_1\lambda_2\dots\lambda_{n-1}}{a_1a_2\dots a_n}\,, \\
&&(A\tilde p,\tilde p) = -4\sum\limits_{k=1}^{n-1} \frac{\left(
\lambda_{k}-a_{1}\right) \cdots \left( \lambda _{k}-a_{n}\right)
\lambda _{k}}{\prod_{s\neq k}\left( \lambda _{k}-\lambda
_{s}\right) }\mu_{k}^{2}.
\end{eqnarray*}

Therefore, the Hamiltonian (\ref{ham:ves}) has the St\"ackel form
in spheroconical variables  and the geodesic flow on $T^*S^{n-1}$
determined by $\mathcal H$ is completely integrable (see
\cite{FeJo}). We have Poisson commuting, quadratic in momenta
integrals $F_1,\dots,F_{n-1}$ (e.g., see \cite{AKN}). One can
prove that functions $F_i$ commute with $H$ using the direct
calculations in elliptic coordinates.

Alternatively, note the geodesic flow of $\mathcal H$, over a
generic invariant torus $\mathbf T$ (level set of
$F_1,\dots,F_{n-1}$) is quasi-periodic with non-resonant
frequencies (for example this follows from Remark \ref{elipsoid}).
Thus, since $\{\mathcal H,K\}_d=0$ and the integral trajectories
are dense on $\mathbf T$, $K$ is also constant along the
Hamiltonian flows of $F_i$ over $\mathbf T$. Since we deal with
analytic functions, we get that $K$ is in involution with
$F_1,\dots,F_{n-1}$ on the whole $T^*S^{n-1}$ ($K$ is the analogue
of the classical Joachimsthal's integral of the geodesic flow on
the ellipsoid $E^{n-1}$ \cite{Mo}). Further, the Hamiltonian $H$,
as a linear combination of $K$ and $\mathcal H$, Poisson commutes
with $F_i$ as well. Whence, the system (\ref{ham_q}),
(\ref{ham_p})  is completely integrable on $T^*S^{n-1}$.

The last assertion of the Theorem follows from the
Liouville-Arnold theorem \cite{AKN} and the fact that the systems
transform to a Hamiltonian form after the same time
reparametrization \eqref{CM}.
\end{proof}

The system is integrable even if not all $a_i$ are distinct. For
any pair of equal parameters $a_i=a_j$, the geodesic flow
(\ref{ham_q}), (\ref{ham_p}) has the additional linear integral $
f_{ij}=\gamma_i \tilde p_j - \gamma_j \tilde p_i. $ For example,
let $n=4$ and $a_1=a_2\ne a_3=a_4$. Then the complete set of
commuting integrals is $f_{12}$, $f_{34}$ and $H$. If we have at
least three equal parameters, the system is integrable according
to the non-commutative version of the Liouville theorem.

\begin{remark}{\rm
Note that the operators \eqref{operator} and
\eqref{inertia_tensor} are related via
$$
D\mathcal I=\mathbf I + DI^{-1}  \quad \Longleftrightarrow \quad
I=D(D\mathcal I-\mathbf I)^{-1}.
$$
In order to reobtain Fedorov's correspondence \eqref{tr1} for
$n=3$ and $f_1=F_1=0$, instead of \eqref{inertia_tensor} one
should consider the inertia operator multiplied by $D$.
}\end{remark}

\subsection{Lagrange Case}
Consider the Lagrange case \eqref{sing}. Due to the additional
$SO(n-1)$-symmetry, the geodesic flow (\ref{ham_q}), (\ref{ham_p})
has the integrals $f_{ij}$, $1\le i<j \le n-1$. Thus, in the
original coordinates we get integrals
\begin{equation}
\label{SI} F_{ij}=(\gamma,A^{-1}\gamma)(\gamma_i p_j - \gamma_j
p_i)^2, \qquad 1\le i<j \le n-1.
\end{equation}

In this case we do not need Hamiltonization to integrate the
reduced system, it is already integrable according to the {\it
Euler-Jacobi theorem}.  Since the generic invariant manifolds
given by $H$ and integrals (\ref{SI}) are two-dimensional and the
system has an invariant measure we have \cite{AKN}:

\begin{thm}
The Lagrange case of the reduced Chaplygin system \eqref{ro1},
\eqref{ro2} is solvable by quadratures; compact regular invariant
manifolds given by functions \eqref{SI} and \eqref{hamiltonian*}
are two-dimensional tori.
\end{thm}

\begin{remark}{\rm
Although the Lagrangian (\ref{ch-lagr}) is additionally invariant
with respect to the right $SO(n-1)$-action, the integrals
(\ref{SI}) are not Noether's integrals. The reason is that the
associated vector fields do not satisfy constraints
(\ref{ch-constr}). For $n=3$ and $I_1=I_2$, the corresponding
integral of the system (\ref{Chap}) is
$F=k_3^2-D(\gamma,(I+D\mathbf I)^{-1}\gamma )k_3^2$.}\end{remark}

\subsection*{Acknowledgments} I am greatly thankful to Yuri N. Fedorov for useful discussions.
I would also use the opportunity to thank the referees for their
detail reports,  valuable remarks and comments which helped me to
correct misprints and to improve the exposition. Remark
\ref{referee} is pointed out by the first referee.

The research was supported by the Serbian Ministry of Science,
Project 144014, Geometry and Topology of Manifolds and Integrable
Dynamical Systems.

\end{document}